# Microdroplet target synthesis for kilohertz ultrafast lasers

Pavel Chvykov[a)], Wise Ongg, James Easter, Bixue Hou, John Nees, and Karl Krushelnick

*Center for Ultrafast Optical Science, College of Engineering, University of Michigan, Ann Arbor, MI 48105 USA*
[a)]primary contact: pchvykov@umich.edu

We have developed a method for producing spatially stable micron-scale liquid targets of flexible shapes at kHz repetition rate for use in air and vacuum, by perturbing 5 and 30-μm diameter streams with fs laser pulses and monitoring the temporal development of the perturbation. Using water, we have produced features such as 2.1-μm diameter droplet and 1.3-μm diameter neck with less than ±0.3-μm shot-to-shot variation, with prospects for further reduction in size and variability. The use of such micron-scale targets can be expected to prevent conductive heat dissipation, enhance field strength for ion acceleration and allow spatially-deterministic laser-cluster experiments.

## 1. Introduction

Recent progress in the field of ultrafast laser science has allowed for many new experiments in the ultra-relativistic regime with exciting applications. These include fast ignition of inertial confinement nuclear fusion[1], as well as more versatile proton acceleration[2], which can be used for cancer therapy. However, most practical applications, in addition to high intensities, also require high repetition rates (up to kHz) – for example to deliver the necessary dose of radiation to kill a tumor. This provides additional difficulties for the set up of the laser and the experiment. One major problem associated with high rep-rate experiments is obtaining a suitable new target for each shot, while maintaining shot-to-shot spatial stability. Thus, solid density target experiments are scarcely done with high rep-rate lasers. Some common solid targets presently used include a rotating wheel that is scanned by the laser focus similar to a compact disc[3], or a metallic tape no less than a few μm thick[4]. Since these targets have to be mechanically translated throughout the experiment, they are limited in stability (typically ± a micron or more) and extensive in size, while the applications such as those described above require single-micron size solid density targets and a submicron precision between shots.

## 2. Overview

In this work, we developed a method for producing liquid targets by perturbing a 5 μm and 30 μm diameter streams with femtosecond laser pulses. Based on its specifics, the perturbation subsequently develops into features of various shapes and sizes. Previously, one paper studied similar perturbation dynamics in a water droplet, but on much larger size scales (~100 μm) and with no intentions or possibility of application to high rep-rate target synthesis[5]. Some of the advantages of the produced targets over what is presently used are that, first, the target is not translated by any mechanical means, and thus the purity of water and the stability of the perturbing laser define the upper limit on the stability of the resulting feature, allowing for much more stable targets. Second, this method allows the formation of targets with a wide range of shapes and sizes, since these can be easily controlled by varying the position and intensity of the perturbing pulse. As a poof of principle, using water, we produced an isolated droplet of 2.1 μm diameter, a cylindrical neck of 1.3 μm diameter and a flat sheet of 1.6 μm thickness with better shot-to-shot spatial stabilities than the resolution limit of our imaging system: less than ±0.3 μm. Such size scales and stabilities are already several times better than what is presently available at high rep-rates, and we can expect that by improving the stability of the perturbing pulse and making its focus tighter (via higher f-number focusing optics and shorter wavelengths from harmonics of the 800 nm used here), these results could potentially be improved.

## 3. Benefits of a micron spherical target

Of particular interest is the 2 μm droplet that was produced – the smallest that was previously published for use in vacuum was 10 μm[6,7] – more than two orders larger in volume, and with much less flexibility than that of the method presented here. It has been experimentally shown that droplets 10-25 μm in diameter are effective for fusion[8] and for generation of ion beams[9] and x-rays[6], and there are multiple reasons why further reduction in size is advantageous. First, because there is no substrate, any conductive heat dissipation is blocked, whereas in conventional macroscopic targets it has been experimentally shown that the incident energy may spread over the surface to more than 100 times the area of the original interaction region within ~500 fs of the laser shot[10], a timescale typical of fusion[8] or laser-cluster interactions[11]. Thus, for micron scale targets, the target is heated entirely, and so the temperature scales inversely with the volume of the droplet. Second, since the lower intensity spatial wings of the beam do not interact with a single-micron target, the effective spatial contrast is improved, thus avoiding the plasma from the interaction of the wings expanding into the path of the beam's center[12]. Third, at relativistic laser intensities, a smaller target allows for higher electron densities at the target's back surface, which enhances field strength for ion acceleration[2,9]. And finally, the size of this droplet approaches the scale of clusters, thus potentially

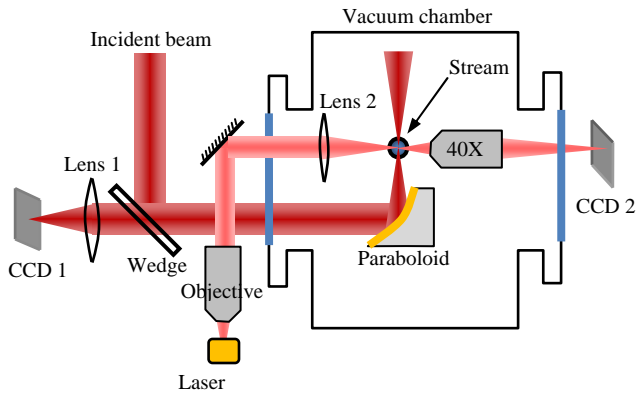

FIG. 1. Experimental Setup

allowing spatially-deterministic laser-cluster experiments, which have not been done to date[11].

## 4. Experiment

The experimental setup used to generate and measure the water micro-targets is shown in figure 1 (see below for technical details). Here, after reflection from a wedge, the main laser pulse enters the vacuum chamber, where it is focused by an ~f/2.5 paraboloid onto a 5 or 30 µm diameter stream, introducing a tiny perturbation. The paraboloid along with lens 1 then images the retro-reflection of the perturbing pulse from the stream onto CCD 1, thus allowing alignment of the focal position across the stream.

For the purpose of imaging the interaction, a laser diode triggered from the perturbing laser with variable delay is focused onto the interaction zone to provide illumination, which is then imaged by an objective onto CCD 2 (see Fig. 2). This allows time-resolved observation of the development of the induced perturbation. At various delays and intensities, the perturbations evolve into structures such as those shown in figure 2. Note that since the diode was pulsed at 500 Hz, while the CCD only took 30 frames per second, each of these images is the integration of ~17 distinct exposures/shots, and thus the contrast of the features can also provide a rough estimate for their stability.

Following are some more technical details of the experiment. The perturbing laser used was a Ti:Sapphire chirped-pulse-amplified laser with 800 nm central wavelength, 500 Hz repetition rate, up to 3mJ pulse energy, and ~30 fs pulse duration. This pulse was then focused via an ~f/2.5 paraboloid to a 2.4±0.3 µm FWHM focus with pointing stability of ±0.3 µm; onto a water stream ejected from a pulled fused silica capillary tip, with a backing pressure of ~5 MPa provided by a compressed gas cylinder. For imaging the interaction, a laser diode was used with 5 mW peak power, driven at 2.3 V with λ=670 nm, pulsed with 100 ns duration. Finally, the CCD-s used were color web-cams with a measured pixel size of 4.1±0.2 µm, which, combined with the magnification optics, provided ~0.1 µm/pixel imaging, which is finer than the resolution allowed by diffraction.

## 5. Results

The photographs in Fig. 2 display the different general types of features that have been observed to develop out of the initial perturbation introduced by the laser, in order of decreasing perturbation intensity: Fig. 2(a), a thin bubble of water formed just 100 ns after the laser shot that explodes the center of a 30 µm stream with intensity on the order of $10^{15}$ W/cm$^2$ (exact value proved to be unimportant) (thickness: 1.6 ±0.3 µm – ray tracing suggested that the visual thickness of the wall is a good estimate for this value, stability: ±0.9 µm); Fig. 2(b), a sideways droplet (the larger one) shaped by surface tension out of a lateral jet driven from the 5 µm stream 750 ns after the initial micro-explosion (droplet diameter: 2.1±0.3 µm, stability: ±0.3 µm, perturbing intensity I≈3×10$^{14}$ W/cm$^2$, stream velocity=29±2 m/s, droplet perpendicular velocity=5.5±0.3 m/s); Fig. 2(c), an inline droplet which forms 6.35 µs after the laser shot as a result of the long-term development of a minor instability introduced to the stream at I≈1×10$^{14}$ W/cm$^2$ (satellite diameter: 3.3±0.3 µm, stability: ±1.0 µm); Fig. 2(d), a cylindrical necking that occurs just before the stream breaks up into droplets – at a delay of 3.15 µs after the perturbation of intensity I≈5×10$^{13}$ W/cm$^2$. This necking is extremely thin and stable, with a diameter of 1.3±0.3 µm and stability of ±0.3 µm (both these uncertainties are limited from below by the resolution of the imaging system). As can be seen from comparing these four scenarios, the observed features change dramatically with time delay and laser intensity, showing the flexibility of this method. Note that due to nearly identical properties of water and heavy-water, we expect all these results to hold equally for both cases.

## 6. Discussion

Several points still remain that require consideration. Firstly, all the above experiments have been carried out both in air and in low vacuum (5.8±0.5 mbar), yielding similar dynamics in both cases, which is to be expected as the laser intensity used is only slightly above the air ionization threshold (~10$^{14}$ W/cm$^2$ [13], though slightly better stability was observed in vacuum), and the viscosity of air is nearly constant over this pressure range. Nonetheless, at higher vacuum levels, we can still expect the dynamics to be similar, at least at the early delays (see Fig. 2, a and b), where the explosion shock waves and surface tension have a much stronger effect than air drag – the force due to surface tension has been estimated to be at least 150 times stronger than that due to air drag in atmosphere. Note that it is possible to achieve significantly higher vacuum with this setup since the stream almost instantly friezes, and the sublimation of frozen water is slow enough for most vacuum pumps to accommodate – extensive work has been done in achieving high vacuum (e.g. $10^{-6}$ mbar) with similar setups for photoelectron spectroscopy[14].

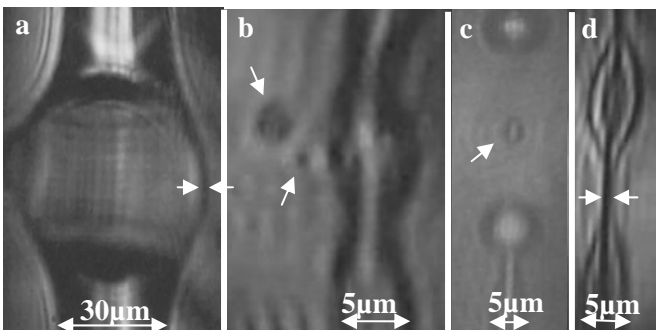

FIG. 2. Various features of 1 µm order in vacuum

Secondly, evaporation and freezing are both common issues related to water in vacuum. When working with higher vacuum, the stream freezes upon hitting the chamber wall, forming an ice pillar that even with a 5 µm horizontal stream can reach a length of over 20 cm in a matter of seconds, potentially freezing over the nozzle. However, this problem has many obvious solutions, e.g. a vibrating catcher. The problem with evaporation arises when the "atmosphere" of vapor around the stream is sufficiently dense for its optical breakdown to distort the main interaction pulse used for experiments. However, from simple kinetic theory considerations at water-vapor equilibrium, it can be shown that the upper limit on the density of vapor immediately above the surface of a 5 µm stream is ~$10^{-5}$ g/cm$^3$, hence the plasma accounts for no more than 0.02×λ phase shift even for intensities as high as I=$10^{23}$ W/cm$^2$, and this shift scales with log(I).

Finally, one of the main limitations on the experiments presented in this paper was the imaging technique used. The use of coherent illumination with wavelength close to 1 µm made it impossible to reliably image features much smaller than 1 µm. Specifically, in figure 2, image (b), we can observe a small droplet-like feature of 0.8 µm diameter to the lower right of the larger droplet; however, there is no way to reliably differentiate this from an interference artifact, and thus no conclusions can be made. Additionally, noting that the recorded droplet diameter of 2.1 µm is nearly the same as the focal spot size, and thus the smallest that could be expected with this method, leads us to believe that a tighter focus could result in smaller features. Hence, using shorter wavelengths for imaging and decreasing the focal spot size may allow for other promising behaviors and features to be observed.

## 7. Conclusion

In this work, we developed a method for producing highly spatially stable liquid targets of flexible shapes and sizes for use with high rep-rate (e.g. kHz) laser systems. This was done by perturbing a 30 and a 5 µm stream with a fs laser pulse and monitoring the temporal development of the perturbation. As proof of principle, we have shown experimentally that with water, this method can produce features of single-micron dimensionality (e.g. a 2.1 µm diameter droplet) and shot-to-shot stabilities of better than ±0.3 µm, where both these results are already several times better than what is presently available. We can also expect that by using a tighter focus for the perturbing pulse and improving its pointing stability, this method can yield even smaller features with higher stabilities.

## 8. Acknowledgement

The authors acknowledge support from the National Science Foundation, the US Naval Research Laboratory, and DTRA.